\documentclass{article}
\usepackage{spconf,amsmath,amssymb,booktabs,graphicx,hyperref}
\hypersetup{hidelinks}

\newcommand\blfootnote[1]{
    \begingroup
    \renewcommand\thefootnote{}\footnote{#1}
    \addtocounter{footnote}{-1}
    \endgroup
}

\newcommand{\best}[1]{\textbf{#1}}
\newcommand{\secondbest}[1]{\underline{#1}}
\newcommand{\primarymodel}{MN-P}

\title{What survives the codec shift: Pooled no-vocals residuals for speech deepfake detection}
\twoauthors
{Jiajun Xu\textsuperscript{*}}
{
Department of Engineering Physics\\
Tsinghua University\\
Beijing, China\\
jiajunx.cv@gmail.com
}
{Menglu Li\textsuperscript{*}, Xiao-Ping Zhang\textsuperscript{$\dagger$}}
{
Tsinghua Shenzhen International Graduate School\\
Tsinghua University\\
Shenzhen, China\\
menglu.li@sz.tsinghua.edu.cn, xpzhang@ieee.org
}

\begin{document}
\ninept
\maketitle
\begin{abstract}
The transition from vocoder-based to neural-codec speech synthesis makes generalization more difficult for speech deepfake detectors, particularly those relying on speech-oriented representations. It remains unclear which acoustic representations retain discriminative information when the generation mechanism changes. We therefore compare 12 acoustic representations using a shared low-capacity linear classifier to identify effective evidence under codec shift. The analysis shows that hierarchical XLS-R leads on the pooled test set, while pooled no-vocals residual statistics perform best on the unseen-codec condition, revealing complementary behavior across generation conditions. Building on this finding, we propose MN-P, a dual-view detector that integrates an utterance-level pooled no-vocals representation with token-level XLS-R features through adaptive gating. The proposed MN-P reduces EER by 54.2\% overall and by 60.9\% on the codec-unseen condition relative to the best-performing retrained state-of-the-art system, with consistent gains across different detector backends. These results indicate that pooled no-vocals residual statistics provide effective complementary evidence for cross-generation speech deepfake detection.

\end{abstract}
\section{Introduction}
\label{sec:introduction}

Speech deepfake detection was initially developed mainly against vocoder-based text-to-speech and voice-conversion systems \cite{tak2021rawnet2,10337724}. With the rapid development of neural audio codecs, codec-based synthesis has become an increasingly important generation paradigm \cite{wu2024codecfake,chen2025neural}. Its high-fidelity outputs and substantially different synthesis mechanism make unseen codec-based attacks a difficult generalization setting. 
\blfootnote{* These authors contributed equally.} 
\blfootnote{$\dagger$ Corresponding author: Xiao-Ping Zhang} 

\begin{figure}[t]
  \centering
  \includegraphics[
    width=\columnwidth
  ]{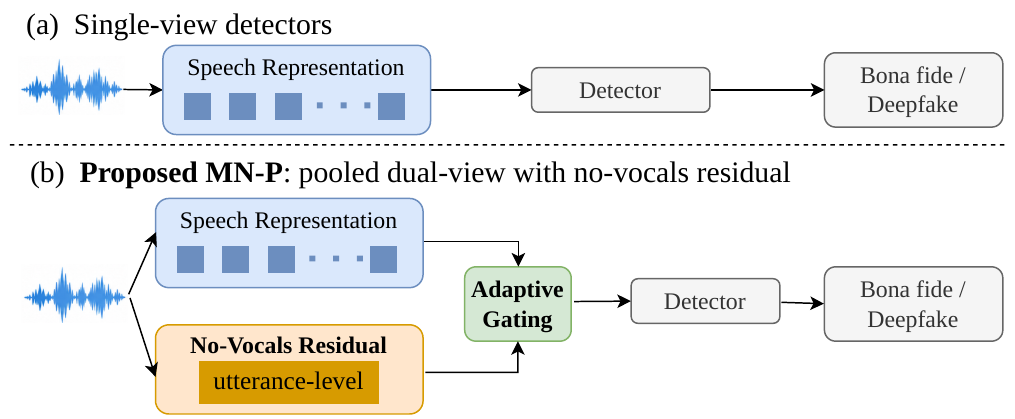}
  \caption{Speech-centric single-view detection versus the proposed MN-P.
(a) A conventional detector relies on a single speech representation.
(b) The proposed MN-P detector complements token-level speech features with an utterance-level
pooled no-vocals residual obtained after vocal separation, and integrates
the two through adaptive gating.}
  \label{fig:method-comparison}
\end{figure}

Most detectors derive their evidence from the speech-bearing signal, using either handcrafted spectral and phase cues \cite{li2022comparative,Wang2023prodoscic} or pretrained self-supervised representations \cite{pascu2024generalizable,zhang2024xlsrsls,li24oa_interspeech}. As codec-based synthesis narrows the perceptual gap to real speech, that evidence becomes harder to separate. Improving the speech representation alone may therefore be insufficient, and robust detection may require complementary evidence beyond the dominant speech content that remains informative across generation shifts. 

We therefore begin with a controlled analysis of 12 acoustic representations spanning spectral, codec-residual, non-vocal, and hierarchical self-supervised features, all evaluated with the same low-capacity linear classifier. The analysis reveals a key representation pattern: hierarchical XLS-R provides the strongest overall discrimination, whereas the pooled no-vocals residual is most effective when the codec is unseen and carries less linearly decodable source identity. This finding shifts the design focus from further strengthening the speech representation to exploiting complementary evidence at a different representational granularity, directly motivating the dual-view formulation in Fig.~\ref{fig:method-comparison}. 

Building on this observation, we develop a dual-view detector, named MN-P, in which a  \textbf{M}ixture-of-Experts Fusion (MoEF)-based speech branch is complemented by a \textbf{N}o-vocals view represented in \textbf{P}ooled form. The speech branch uses a frozen self-supervised XLS-R encoder and fuses its hierarchical hidden states through MoEF. In contrast to treating residual information as a parallel frame-level stream, the auxiliary branch summarizes the no-vocals residual into a single utterance-level vector. Adaptive gating then integrates this global residual context with the token-level speech representation without requiring temporal alignment between the two views. This asymmetric design preserves local temporal evidence in the speech representation while introducing utterance-level residual statistics as complementary context. MN-P outperforms retrained state-of-the-art systems on a test set spanning unseen vocoders and an unseen neural codec, with consistent gains across detector backends. Together, these results establish pooled no-vocals residual statistics as an effective complementary representation for cross-generation speech deepfake detection.

\section{Related Work}
\label{sec:related-work}
Speech deepfake detection has evolved from handcrafted spectral and phase cues \cite{sahidullah2015comparison,todisco2017cqcc} to raw-waveform, graph-based, and self-supervised detectors \cite{pascu2024generalizable,jung2022aasist}, with recent work further exploiting intermediate self-supervised learning layers \cite{zhang2024xlsrsls,wang2025moef}. Most systems nevertheless remain speech-centric, deriving detection evidence primarily from speech-oriented representations, which leaves open whether such evidence alone remains sufficient when the synthesis mechanism changes substantially. 

Complementary evidence outside the dominant verbal content has also been explored, including silence, breathing, background, and environmental components \cite{muller2021silence,zhang2021silence,doan2023btse, li2023robust,salvi2024listening,zhang2026compspoof}. However, these cues have mainly been studied without explicitly considering the vocoder-to-neural-codec shift, so their robustness under codec shift remains unclear. 

Generalization to unseen datasets and generators remains challenging \cite{paul2017generalization,muller2022generalize,das2020generalized, li2024crossdomain}. CodecFake and subsequent codec-aware methods address the vocoder-to-neural-codec shift mainly through new training data and learning objectives \cite{wu2024codecfake,xie2025codecfake}, leaving the representation question less explored. 

These limitations motivate us to move beyond strengthening the speech representation alone and identify complementary evidence that remains informative under generation shift. We compare multiple acoustic representations and find pooled no-vocals residual statistics to be the most effective under the unseen-codec condition while complementing hierarchical XLS-R. Building on this finding, we propose MN-P, which integrates an utterance-level pooled no-vocals representation with token-level XLS-R features through adaptive gating. Rather than modeling the residual as another frame-level stream, this formulation aggregates residual statistics that remain stable over the entire utterance.

\section{Representation Analysis}
\label{sec:analysis}
We perform a representation analysis to identify which acoustic representations remain informative when the synthesis condition shifts from vocoder-based to neural-codec-based generation. To isolate representation quality from classifier capacity, we evaluate 12 candidate representations under the same data protocol and a common low-capacity linear classifier, reporting both overall detection performance and performance on codec-unseen (CU) samples.

The candidates span six feature families. B1 uses LFCC \cite{sahidullah2015comparison}, B3 uses phase and group-delay cues, B5 uses EnCodec quantization residuals \cite{defossez2023encodec}, and B6 uses hierarchical XLS-R hidden states \cite{babu2022xlsr}, each contributing one candidate. B2 contains four spectrotemporal-statistic representations derived from different waveform views, namely the original signal, the HTDemucs-estimated \cite{rouard2023hybrid} vocals (V) and no-vocals (NV) components, and their cross-view relation (XV). The no-vocals representation is detailed in Section~\ref{sec:b2}. B4 contributes four candidates obtained by WORLD source--filter analysis \cite{morise2016world} from pooled WORLD statistics and cross-view relations across the original, V, and NV views.

Vocoder-based samples are drawn from ASVspoof2019-LA \cite{todisco2019asvspoof}, and codec-based samples from Codecfake \cite{xie2025codecfake}. Following the official corpus splits, the merged protocol contains 273,027 training, 55,269 development, and 190,845 test samples. The training and development sets cover A01--A06 vocoders and F01--F06 codecs, whereas the test set additionally contains unseen vocoders A07--A19, seen codecs F01--F06, and the held-out DAC codec F07. F07 is absent from both training and development and serves as the CU condition. We report EER over the full test set and CU EER over F07 only. Analysis training uses AdamW with batch size 4,096, at most 30 epochs, patience 5, learning rate $10^{-3}$, and weight decay $10^{-4}$. On the same fixed representations, we additionally train a linear probe to predict fake-source identity. The probe is evaluated in a closed-set setting under the same protocol for all candidates, and its macro-F1 is denoted Source F1. It measures how much source-specific information is linearly decodable from each representation. High Source F1 suggests that separability may rest on cues identifying the generating system, which need not transfer to an unseen generator, so lower values are preferable here.

\begin{table}[!t]
  \caption{Six candidate representations with the lowest overall test-set EER. Source F1 is the closed-set macro-F1 of a linear probe for fake-source identity, where lower values indicate less linearly decodable source information. Bold and underlined denote the best and second-best results.}
  \label{tab:screening}
  \centering
  \setlength{\tabcolsep}{2.5pt}
  \renewcommand{\arraystretch}{1}
  \begin{tabular*}{\columnwidth}{@{\extracolsep{\fill}} l c c c@{}}
    \toprule
 Candidates & EER(\%) $\downarrow$ & CU EER(\%) $\downarrow$ & Source F1 $\downarrow$\\
    \midrule
  B6   & $\best{4.843}$        & $\secondbest{9.291}$ & 0.726 \\
  B2-XV & $\secondbest{13.448}$ & 21.935               & 0.656 \\
   B2-NV  & 13.859      & $\best{8.619}$       & $\best{0.539}$ \\
 B2-V   & 25.999                & 41.201               & $\secondbest{0.588}$ \\
B3  & 26.766                & 34.606               & 0.628 \\
B1   & 26.956                & 34.204               & 0.620 \\
    \bottomrule
  \end{tabular*}
\end{table}

Table~\ref{tab:screening} lists the six candidates with the lowest overall EER on the test set. B6 achieves the lowest overall EER at 4.843\%, while B2-NV achieves the lowest CU EER at 8.619\%, compared with 9.291\% for B6. This establishes B2-NV as the strongest screened representation when the generating codec is unseen. The within-B2 comparison makes this contrast direct. Replacing the vocals view with the no-vocals view reduces EER from 25.999\% to 13.859\% and CU EER from 41.201\% to 8.619\%, and B2-XV reaches a comparable overall EER of 13.448\% but a substantially higher CU EER of 21.935\%. The codec-unseen advantage is therefore specific to the no-vocals view rather than a general property of the B2 statistic family. B2-NV also yields the lowest Source F1 at 0.539, against 0.726 for B6, so its codec-unseen margin comes with less rather than more linearly decodable source information. The non-vocal residual thus carries evidence that is distinct from what the speech-bearing representation encodes and that remains accessible when the generating codec is unseen. We therefore use B6 as the primary representation and B2-NV as a complementary residual view in Section~\ref{sec:method}.

\begin{figure*}[t]
  \centering
  \includegraphics[width=1\textwidth]{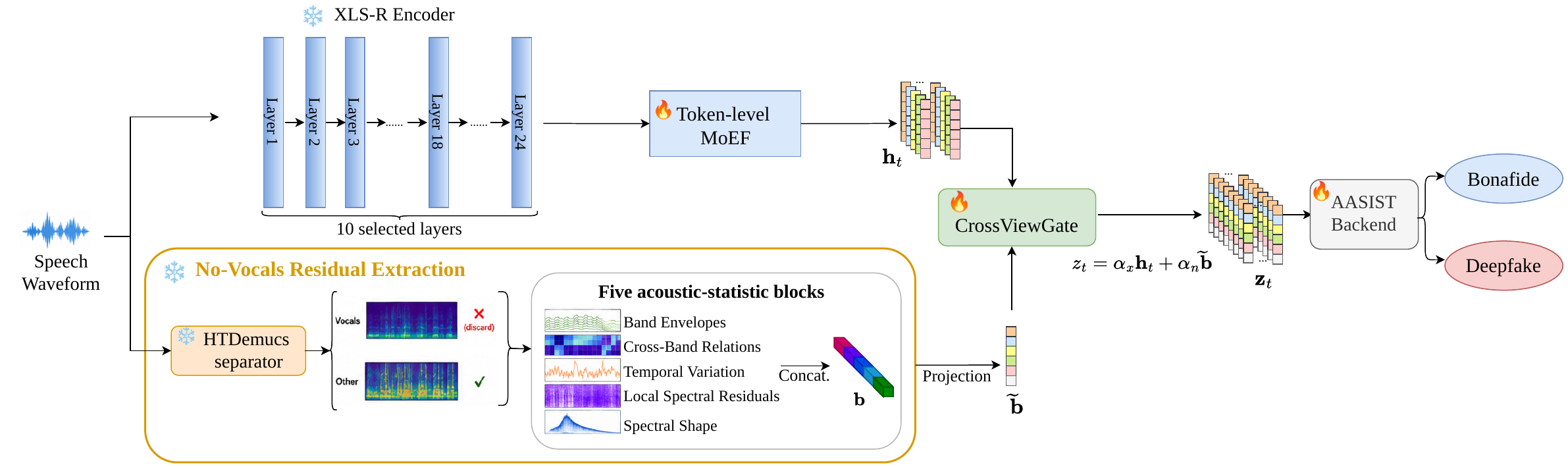}
 \caption{Architecture of the proposed MN-P detector. A frozen XLS-R branch produces a token-level speech representation, while the no-vocals branch, highlighted in the orange box, aggregates residual statistics over the whole utterance into a single vector.  CrossViewGate fuses the two views before the AASIST backend. Frozen and trainable modules are marked by distinct icons. } 
  \label{fig:mn-p}
\end{figure*}

\section{Proposed Method}
\label{sec:method}

Motivated by the complementary behavior identified in Section~\ref{sec:analysis}, we construct MN-P by combining the MoEF-fused hierarchical XLS-R representation with the pooled no-vocals representation. A central design choice is to represent the no-vocals residual as an utterance-level auxiliary vector rather than as a second temporal stream, so that residual statistics are aggregated over the whole utterance without requiring frame-level alignment with the XLS-R sequence. Fig. ~\ref{fig:mn-p} summarizes the architecture.

\subsection{Pooled No-Vocals Residual}
\label{sec:b2}

For a 4-s waveform $x$, HTDemucs \cite{rouard2023hybrid} separates the 44.1-kHz resampled signal into components $D_c(\mathcal{R}_{44.1}(x))$, where $c\in\mathcal{S}=\{\mathrm{vocals},\mathrm{drums},\mathrm{bass},\mathrm{other}\}$. We discard the vocals component, sum the remaining sources, downmix, and resample to 16~kHz: 
\begin{equation} 
\mathbf{n}=\mathcal{R}_{16}\!\left( \operatorname{Downmix}\!\left[ \sum_{c\in\mathcal{S}\setminus\{\mathrm{vocals}\}} D_c(\mathcal{R}_{44.1}(x)) \right]\right). 
\end{equation} 
We refer to $\mathbf{n}$ as the Demucs-estimated no-vocals residual. Rather than treating it as an isolated background source, we use the full residual as a detection representation. It contains background content together with separation artifacts and residual speech energy, all of which can preserve traces introduced by the synthesis pipeline beyond the dominant verbal content. 

We characterize this residual through utterance-level spectrotemporal statistics. A center-free STFT with a 512-point DFT, a 400-sample Hann window, and a 160-sample hop is converted to log power and divided into 0--2, 2--4, and 4--8~kHz bands. Each trajectory is summarized by eight operators: mean, standard deviation, minimum, maximum, the 25th, 50th, and 75th percentiles, and mean absolute first difference. These statistics form five feature blocks capturing band envelopes, cross-band relations, temporal variation, local spectral residuals over a $5\times5$ smoothed surface, and spectral shape. The five blocks contain 24, 35, 27, 24, and 17 dimensions, respectively, and are concatenated as 
\begin{equation} 
\mathbf{b}= [\phi_1(\mathbf{n});\ldots;\phi_5(\mathbf{n})] \in\mathbb{R}^{127}. 
\end{equation} 
The resulting no-vocals residual representation is a single 127-D vector per utterance, rather than a frame-level residual stream, and provides the auxiliary view fused with XLS-R.  Column-wise median imputation and standardization are fitted on the training set only and then fixed for development and test data.

\subsection{Dual-View Integration}
 \label{sec:fusion} 

The primary branch uses the frozen XLS-R-300M encoder \cite{babu2022xlsr}. Following MoEF \cite{wang2025moef}, we select ten hidden states spanning shallow to deep layers and assign four feed-forward experts to each. A shared router uses the layer-24 token to score all 40 experts, and global Top-2 routing selects two layer-expert pairs at each token position. The resulting sequence $\{\mathbf{h}_t\}_{t=1}^{T}$ retains the XLS-R temporal grid with 1024-D features. 

The pooled residual vector is projected to the XLS-R feature dimension by a two-layer network with a 256-D GELU hidden layer and layer normalization, giving $\widetilde{\mathbf{b}}=Project(\mathbf{b})\in\mathbb{R}^{1024}$.  We summarize the XLS-R sequence as $\overline{\mathbf{h}}=T^{-1}\sum_t\mathbf{h}_t$ and use an utterance-conditioned two-view gate, as CrossViewGate, to estimate the relative contribution of the two representations: 
\begin{equation} 
[\alpha_x,\alpha_n]^{\mathsf T} =\operatorname{softmax} G([\overline{\mathbf{h}};\widetilde{\mathbf{b}}]). 
\end{equation} 
The utterance-level residual vector is then broadcast over the XLS-R token grid: 
\begin{equation} \mathbf{z}_t=\alpha_x\mathbf{h}_t+\alpha_n\widetilde{\mathbf{b}}, \qquad t=1,\ldots,T. 
\end{equation} 
Unlike the sparse MoEF router, CrossViewGate is dense over the two views, allowing both to contribute while adapting their relative weights to each utterance. The fused sequence is classified by an AASIST backend \cite{jung2022aasist}.

\begin{table*}[!t]
  \caption{Performance comparison on the merged test set (\%). MN-P is evaluated with AASIST and Nes2Net backends; M0 removes the pooled no-vocals view and CrossViewGate. Selected state-of-the-art systems are retrained under the same protocol. Bold and underlined denote the best and second-best results.}
  \label{tab:main}
  \centering
  \setlength{\tabcolsep}{2.5pt}
  \renewcommand{\arraystretch}{1}
  \begin{tabular*}{\textwidth}{@{\extracolsep{\fill}}lccccc@{}}
    \toprule
    System & AUC $\uparrow$ & EER (\%) $\downarrow$ & Fake miss (\%) $\downarrow$ & CU EER (\%) $\downarrow$ & CU fake miss (\%) $\downarrow$ \\
    \midrule
    \textbf{\primarymodel-AASIST (Ours)}
      & $\best{99.638}$ & $\best{2.744}$ & $\best{6.890}$
      & $\best{6.136}$ & $\best{27.903}$ \\
    \primarymodel-Nes2Net
      & $\secondbest{99.583}$ & $\secondbest{2.927}$ & $\secondbest{9.068}$
      & $\secondbest{6.706}$ & $\secondbest{40.182}$ \\
    M0-AASIST
      & 98.118 & 7.010 & 14.702
      & 18.639 & 68.729 \\
    M0-Nes2Net
      & 98.630 & 5.869 & 14.499
      & 15.632 & 68.092 \\
    \midrule
    FAD~\cite{wang2025moef}
      & 98.507 & 6.246 & 14.876 & 16.562 & 70.012 \\
    W2V2-AASIST with CSAM~\cite{xie2025codecfake}
      & 98.647 & 5.986 & 15.444 & 15.692 & 74.325 \\
    XLS-R-SLS~\cite{zhang2024xlsrsls}
      & 96.068 & 10.716 & 18.228 & 28.180 & 79.290 \\
    W2V2-Nes2Net~\cite{liu2025nes2net}
      & 98.389 & 6.569 & 15.756 & 17.597 & 72.120 \\
    \bottomrule
  \end{tabular*}
\end{table*}

\section{Experimental Setup}
\label{sec:setup}

\subsection{Datasets and Metrics} 
We combine two English corpora. ASVspoof 2019 LA \cite{todisco2019asvspoof} provides vocoder-based fakes and Codecfake \cite{xie2025codecfake} provides neural-codec fakes, with bona fide speech from both. We retain mono 16-kHz, 64,000-sample waveforms with RMS at least $10^{-2}$ and peak amplitude at least $3\times10^{-2}$. 

Training and development use the A01--A06 vocoders and F01--F06 codecs, balanced by corpus and label to 50,758 and 49,678 utterances, respectively. The 39,996-utterance test set is balanced across evaluation conditions, covering bona fide speech, seen and unseen ASVspoof vocoders, the seen codecs F01--F06, the held-out DAC codec F07, and unseen audio language model (ALM)-based deepfake samples from Codecfake \cite{xie2025codecfake}, so that no single condition dominates the evaluation. F07 is excluded from training and development. Its 4,444 fake and 4,444 bona fide utterances define the codec-unseen (CU) subset.

We report AUC, EER, and fake miss on the full test set, and EER and fake miss on the CU subset. AUC and EER are computed from test scores, while fake miss uses the EER threshold selected on the development set.

\subsection{Implementation Details} 
\label{sec:implementation} 
Main-model experiments use seeds 3407, 3411, and 3417, and the reported main-model and ablation results are averaged over the three seeds. Trainable modules are optimized with AdamW using batch size 8, learning rate $10^{-4}$, weight decay $10^{-4}$, at most 10 epochs, and patience 5. The checkpoint with the minimum development loss is used for evaluation. XLS-R remains frozen throughout training, and HTDemucs is applied offline so that the 127-D residual vectors are precomputed once and reused across seeds and systems. All comparison systems share the same data, waveform preprocessing, and evaluation protocol.

\section{Results and Discussion}
\subsection{Main Comparison} 

Table~\ref{tab:main} reports the protocol-matched comparison. The first four rows provide a controlled comparison between MN-P and the corresponding M0 baseline across two detector backends. M0 retains the same hierarchical XLS-R with MoEF branch and detector backend as MN-P, but removes the pooled no-vocals branch and CrossViewGate. With AASIST, MN-P reduces EER from 7.010\% to 2.744\% and CU EER from 18.639\% to 6.136\%. The larger reduction on the unseen-codec subset shows that the pooled residual view is particularly effective under codec shift. The same trend holds with the Nes2Net backend, where CU EER decreases from 15.632\% to 6.706\%. MN-P-Nes2Net ranks second across all reported metrics, indicating that the gain from the pooled no-vocals representation is not specific to a particular detector backend. 

The last four rows in Table~\ref{tab:main} report state-of-the-art detection systems retrained  under the same protocol. Both MN-P variants outperform all retrained systems across the reported metrics, with the largest advantage under codec shift. The strongest of them, W2V2-AASIST with CSAM, reaches 5.986\% EER and 15.692\% CU EER through codec-aware co-training. MN-P-AASIST
reduces these by 54.2\% and 60.9\% relative, without any codec-specific training signal. The difference is even larger in CU miss, where retrained systems miss 70--79\% of codec-unseen fakes, compared with 27.903\% and 40.182\% for the two MN-P variants.

\subsection{Ablation on Model Structure } 

Table~\ref{tab:ablation-form} indicates the contribution of the no-vocals view and its representation form using an independent set of 2,000 training, 1,000 development, and 2,000 test samples over the same three seeds. M0 omits the no-vocals residual view, MN-S retains a frame-level residual sequence, whereas MN-P summarizes the residual as an utterance-level vector. 

The ablation reveals a two-stage gain. Adding the no-vocals sequence reduces CU EER from 34.830\% for M0 to 18.168\%, showing that the residual itself contains useful codec-unseen evidence. Pooling it at utterance level further reduces CU EER to 8.709\% and CU miss from 57.808\% to 14.865\%. This further improvement shows that the no-vocals residual is more effective when summarized at utterance level than when retained as a frame-level sequence, while avoiding explicit alignment with the XLS-R sequence.

\begin{table}[!t]
  \caption{Ablation on model structure (\%), on an independent train/development/test split. M0 uses XLS-R alone and MN-S feeds the no-vocals residual as a frame-level sequence. Bold and underlined denote the best and second-best results.}
  \label{tab:ablation-form}
  \centering
  \setlength{\tabcolsep}{2.5pt}
  \renewcommand{\arraystretch}{1}
  \begin{tabular*}{\columnwidth}{@{\extracolsep{\fill}}lrr@{}}
    \toprule
     & CU EER (\%) $\downarrow$ & CU fake miss (\%) $\downarrow$ \\
    \midrule
    M0 & 34.830 & 80.180 \\
    MN-S & $\secondbest{18.168}$ & $\secondbest{57.808}$ \\
    MN-P & $\best{8.709}$ & $\best{14.865}$ \\
    \bottomrule
  \end{tabular*}
\end{table}

\section{Conclusion}
\label{sec:conclusion}
The shift from vocoder-based to neural-codec speech synthesis makes generalization increasingly difficult for detectors that rely mainly on speech-oriented representations. In this work, we first compared multiple acoustic representations under a shared protocol and found that the representation performing best overall is not necessarily the most effective under unseen-codec conditions. In particular, pooled no-vocals residual statistics outperform hierarchical XLS-R in that condition. Based on this finding, we proposed MN-P, which combines token-level XLS-R features with an utterance-level pooled no-vocals representation through adaptive gating. 

Experimental results show that MN-P more than halves the codec-unseen EER of the best-performing retrained state-of-the-art system, with the gain retained across different detector backends.  A structural ablation further demonstrates that the no-vocals residual is more effective when represented by utterance-level pooled statistics than as a frame-level stream. Together, these results indicate that complementary evidence beyond the speech representation can provide substantial generalization gains under changes in the generation mechanism. Future work will disentangle the contributions of background content and separation artifacts within the residual, and extend the evaluation to more diverse conditions.

\vfill\pagebreak

\end{document}